\documentclass[aps,pre,twocolumn,showpacs,floatfix ]{revtex4}
\usepackage{amsmath,amsthm,amsfonts,amssymb,graphicx}

\begin{document}

\title{Collective Synchronization Induced by Epidemic Dynamics\\
on Complex Networks with Communities}

\author{Gang Yan$^{1}$}
\author{Zhong-Qian Fu$^{1}$}\email{zqfu@ustc.edu.cn}
\author{Jie Ren$^{2}$}
\author{Wen-Xu Wang$^{2}$}
\affiliation{$^{1}$Department of
Electronic Science and Technology, \\
University of Science and Technology of China, Hefei, Anhui, 230026,
P.R.China\\
$^{2}$Nonlinear Science Center and Department of Modern Physics, \\
University of
Science and Technology of China, Hefei, Anhui, 230026, P.R.China}

\date{\today}

\begin{abstract}
Much recent empirical evidence shows that \textit{community
structure} is ubiquitous in the real-world networks. In this
Letter, we propose a growth model to create scale-free networks
with the tunable strength (noted by $Q$) of community structure
and investigate the influence of community strength upon the
collective synchronization induced by SIRS epidemiological
process. Global and local synchronizability of the system is
studied by means of an order parameter and the relevant
finite-size scaling analysis is provided. The numerical results
show that, a phase transition occurs at $Q_c\simeq0.835$ from
global synchronization to desynchronization and the local
synchronization is weakened in a range of intermediately large
$Q$. Moreover, we study the impact of mean degree $\langle
k\rangle$ upon synchronization on scale-free networks.
\end{abstract}

 \maketitle

\emph{I. Introduction.}---The study of networked systems, including
technological, social and biological networks of various kinds, has
attracted much attention in physics community
\cite{albert-review,dorogovtsev-review,newman-review,Pastor-book}.
How the properties of networks, such as the lengths of shortest
paths between vertices, degree distribution, clustering coefficient,
degree-degree correlation and so on, affect dynamical processes
taking place upon the networks
\cite{Pastor1,Pastor2,Lai,corr,bete,deg}, has been one of the most
important subjects of the body of work. Recently, it has been
determined that many real-world networks show \textit{community
structure} \cite{Newman1,Vicsek}, i.e., groups of vertices that have
a high density of edges within them, while a lower density of edges
between groups. However, there's few work about the influences of
various degree of community structure upon dynamics.

In this paper, we intend to fill this gap by investigating
synchronization behavior induced by the SIRS epidemiological
dynamics \cite{SIRSmodel1,SIRSmodel2} on the scale-free networks
with various strength (noted by $Q$) of community structure. In Ref.
\cite{SIRS}, the authors have studied the SIRS on small-world
networks and found that when $p$, which characterizes the degree of
disorder of the network, reach an intermediately large value $p_c$,
synchronization of the system emerges. Comparatively, we focus on
global and local (inside each community) dynamics, and discover that
no synchronization comes forth when the network possesses strong
enough community structure, i.e. the communities are connected by
few edges among them. Moreover, the vertices inside each community
behave weaker synchronization when $Q$ is in a range of
intermediately large values.

\emph{II. Network Model.}---To generally characterize the community
structure of scale-free networks, we propose a growth model to
create a network with a tunable parameter denoting the strength of
community structure. Inspired by two ingredients of Barabasi-Albert
model (BA for short), i.e., growth and preferential attachment
\cite{BAmodel}, the rules of our model are as follows: Starting with
$c$ communities, noted by $U_1,U_2,...,U_{c-1},U_c$, and each
community with a small number ($m_0$) of vertices. At every time
step, we add into each community a new vertex with $m$($<m_0$) edges
that link the new vertex to $n$ different vertices in this community
and $m-n$ different vertices in other $c-1$ communities already
existed in the system. The initial $m_0\times c$ vertices link to
each other to keep the connectivity of the network. The values of
$m$ and $n$ are not necessary integers (take $m$ for example: the
fractional part of $m$ denotes the probability to link $m^\prime+1$
different vertices, where $m^\prime$ is the integral part of $m$).
When adding a new vertex into community $U_l$, firstly choose $n$
different vertices in community $U_l$ according to ``preferential
attachment", which means the probability $\prod$ that the new vertex
will connect to vertex $i$ ($i\in U_l$) depends on the degree $k_i$
of vertex $i$, such that $\prod(k_i)=k_i/\sum_{j\in U_l}k_j$. Then
for each one of the other $m-n$ edges of the new vertex, choose a
community $U_h$($\neq U_l$) randomly and connect the new vertex to
one vertex in $U_h$ following the preferential attachment mechanism
referred above.

The scaling behavior of the degree distribution can be calculated by
using several approaches \cite{continuum,master,rate}. In our model,
the degree distributions $p(k)$ of vertices of the global network,
as well as the local vertices (inside each community), are power-law
with exponent $3.0$, i.e., $p(k)\propto k^{-3.0}$ (see Fig.1). The
analytic procedure is simple and not shown here.

\begin{figure}[tp]
\centering
%\leavemode
\includegraphics[width=\columnwidth ]{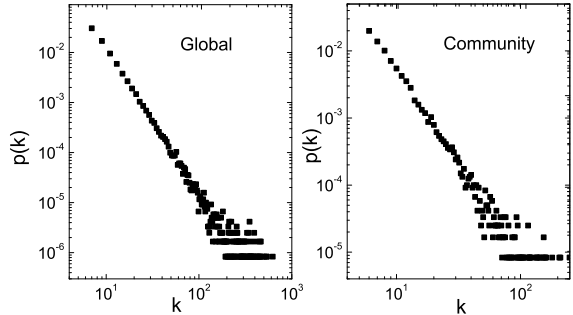}
\caption{The global (left) and local (right) degree distribution of
the network with $N=10^5$, $c=10$, $m=4.0$ and $n=3.0$, that says
$Q=0.65$. It is worthwhile to point out that, for different values
of $Q$, the distributions do not change.}
%\end{center}
\label{fig1}
\end{figure}

As proposed by Newman and Girvan \cite{q-value} and modified by
Kashtan and Alon \cite{Alon}, the strength of community structure
can be quantified by
\begin{equation}
Q=\sum^{c}_{1}\bigg[\frac{l_s}{L}-\bigg(\frac{d_s}{2L}\bigg)^2\bigg],
\label{eq.1}
\end{equation}
where $c$ is the number of communities, $L$ is the number of edges
in the network, $l_s$ is the number of edges between nodes in
community $U_s$, and $d_s$ is the sum of the degrees of the nodes in
community $U_s$. Roughly speaking, $Q$ is the ratio of the number of
edges intra-community to the total number of the edges. Obviously,
if the network is divided into some communities more clearly, i.e.
there are fewer edges among different communities, the value of $Q$
is larger. In our model, for large N (the number of all vertices),
$L=mN$, $l_s=\frac{nN}{c}$ and
$d_s=2n+(m-n)+(m-n)*(c-1)/(c-1)=\frac{2mN}{c}$. Substituting these
results into Eq.\ (\ref{eq.1}), we obtain
\begin{equation}
Q=\frac{n}{m}-\frac{1}{c}. \label{eq.2}
\end{equation}
Thus, for fixed $m$ and $c$, we modulate the value of $n$ to get
the networks with various community strength $Q$.

\emph{III. Epidemic Model.}---We analyze SIRS epidemic model and
aim to point out the role of community structure on the temporal
dynamics of the epidemic spreading. The disease has three stages:
susceptible (S), infected (I), and refractory (R). A vertex of the
networked population is described by a single dynamical variable
adopting one of these three values. Susceptible elements can pass
to the infected state through contagion by an infected one.
Infected elements pass to the refractory state after an infection
time $T_I$ . Refractory elements return to the susceptible state
after a recovery time $T_R$. The contagion is possible only during
the S phase, and only by an I element. During the R phase, the
elements are immune and do not infect. The system evolves with
discrete time steps. Each vertex in the network is characterized
by a time counter $\tau_i(t)=0,1,...,T_I+T_R\equiv T$, describing
its phase in the cycle of the disease. The epidemiological state
$\pi_i$ (S, I, or R) of the vertex depends on the phase in the
following way:
\begin{equation}
\begin{array}{ll}
\pi_i(t)=S & \;\;\mbox{if}~\tau_i(t) = 0 \\
\pi_i(t)=I & \;\;\mbox{if}~\tau_i(t) \in [1,T_I] \\
\pi_i(t)=R & \;\;\mbox{if}~\tau_i(t) \in [T_I\!+\!1,T]
\end{array}\label{eq.3}
\end{equation}

\begin{figure}[tp]
\centering
%\leavemode
\includegraphics[width=\columnwidth ]{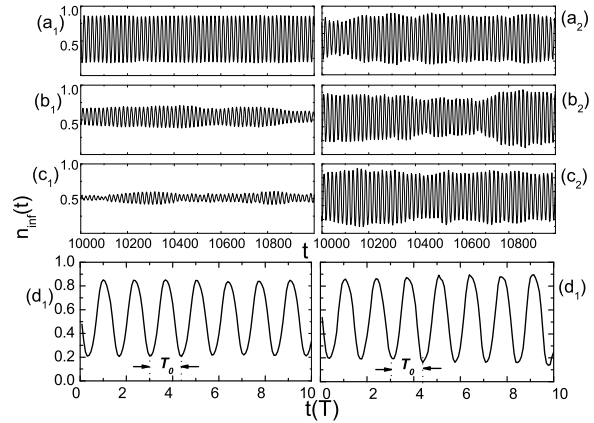}
\caption{The time series of the fraction of infected vertices. The
systems have $N=10^4, c=25$ and $m=4.0$, and the infection cycle
with $T_I=8$ and $T_R=5$. The left three figures (a1), (b1) and (c1)
showed the global fluctuations of $n_\text{inf}(t)$ on the network
with $Q=0.46, 0.81$ and $0.935$ respectively. The right three
figures showed the local fluctuations correspondingly. It's obvious
that the global and local fluctuations are very different. The
detailed analysis is presented in the text. The two bottom figures
show the clear global (d1) and local (d2) periodic oscillations on
the network with weak community structure ($Q=0.46$). The time steps
have been scaled by the natural period $T$ of the infection cycle.
$T_0$ is the period of the oscillations. It is manifest that
$T_0>T$, which is different from the result $T_0=T$ presented for SW
networks in Ref.\cite{Gade}.}
%\end{center}
\label{fig1}
\end{figure}

The state of a vertex in the next step depends on its current
phase in the cycle, and the state of its neighbors in the network.
A susceptible vertex stays as such, at $\tau=0$, until it becomes
infected. Once infected, it goes (deterministically) over a cycle
that lasts $T$ time steps. During the first $T_I$ time steps, it
is infected and can potentially transmit the disease to a
susceptible neighbor. During the last $T_R$ time steps of the
cycle, it remains in state R, immune and not contagious. After the
cycle is complete, it returns to the susceptible state. As
mentioned in Ref. \cite{SIRS}, if vertex $i$ is susceptible and it
has $k_i$ neighbors, of which $k_{\text{inf}}$ are infected, then,
$i$ will become infected with probability $k_{\text{inf}}/k_i$.

\emph{IV. Results and Analysis.}---Specifically we study the
behavior of the infected sites with respect to $Q$. A typical
realization starts with the generation of the network
characterized by $Q$ and the initialization of the states of the
vertices. The initial fraction of infected vertices
$n_{\text{inf}}(0)=0.1$ and the rest susceptible, was used in all
the simulations here.

\begin{figure}[tp]
\includegraphics[width=\columnwidth]{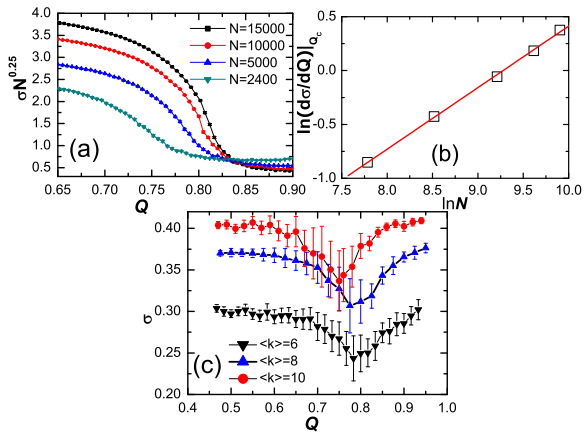}
\caption{(Color online) (a) shows the order parameter for global
synchronization plotted as $\sigma N^{\beta/\overline{\nu}}$ with
$\beta/\overline{\nu}=0.25$ vs $Q$ for different network size $N$
with fixed $Nc=200$, where $Nc$ is the number of vertices in each
community. There is given a unique crossing point at $Q_c=0.83(5)$.
From (b) we obtained $(1-\beta)/\overline{\nu}=0.57(2)$. These yield
$\beta\approx0.30$ and $\overline{\nu}\approx1.22$. (c) displays the
order parameter for local synchronization vs $Q$ for different
network mean degree $\langle k\rangle=6$, 8, 10 (from bottom to
top).} \label{fig3}
\end{figure}

\begin{figure}[tp]
\centering
%\leavemode
\includegraphics[width=\columnwidth ]{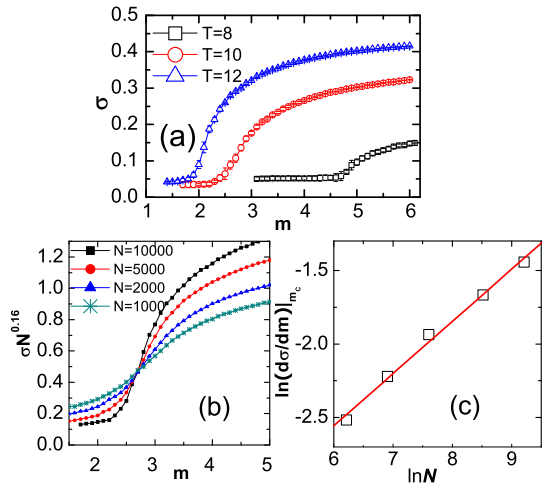}
\caption{(Color online) (a) shows the order parameter vs $m$ for
different natural period $T=8$, 10, 12, where $m=\langle k
\rangle/2$. (b) displays the order parameter $\sigma$ for natural
period $T=10$ plotted as $\sigma N^{\beta/\overline{\nu}}$ with
$\beta/\overline{\nu}=0.16$ vs $m$ for different network size $N$.
There is given a unique crossing point at $m_c=2.80$. From (c) we
obtained $(1-\beta)/\overline{\nu}=0.35(0)$. These yield
$\beta\approx0.31$ and $\overline{\nu}\approx1.95$.}
%\end{center}
\label{fig1}

\end{figure}
After a transient period, a stationary state is achieved. We find
that the pronounced fluctuations of the fraction of infected
vertices is a function of time. Figure 2 shows three time series
displaying the fraction of infected vertices in the network with
varying community strength $Q$. When $Q=0.46$ (see Fig.2(a1)), the
network has a weak strength of community structure. It is similar to
the real-world networks where the community strength $Q$ falls in
the range from about 0.3 to 0.7 \cite{q-value}. In such condition,
the fraction of infected vertices exhibits large amplitude
oscillations. For strong community structure, such as $Q=0.81,
0.935$ (see Fig.2(b1) and (c1) respectively), the time series have
regular periods but the amplitudes is small and disordered. In
addition, we study the local dynamics, that is the epidemic process
inside each community. Figure 2(a2), (b2) and (c2) show the time
evolution of the fraction of infected vertices in a community, for
$Q=0.46, 0.81$ and 0.935 respectively. The amplitudes are almost the
same. Since the amplitude is related to the synchronization of the
system, we will give a measure below and make it clearer. Fig. 2(d1)
and (d2) show the clear periodic oscillations of the fraction of
local and global infected vertices vs the scaled time $t/T$ while
$Q=0.46$, respectively. One can see that the period $T_0$ is larger
than the natural period $T$ of the infection cycle, which is
different from the result on small-world networks presented in Ref.
\cite{Gade}. Moreover, we have done the Fourier power analysis for
different $Q$ and find that there is a sharp peak in the frequency
$1/T_0$ (not showed here), which reveals that the series have
regular temporal periods although the amplitudes are variable. As
the difference between $T_0$ and $T$ is the time staying at the
state S and $T$ is of the same value for all vertices, we could
analyze the reason of regular $T_0$ by estimating the probability
$P_{i}(t)$ of a vertex $i$ changing state from S to I at time $t$.
We here obtain $p_i(t)$ by using the mean-field estimation as the
following, $p_i(t)=k_{inf}/k_i\propto n_{inf}(t)$. That results
implies that all the vertices update their states at almost the same
time which induces the regular temporal cycles. Besides the
parameter-free infection mechanism, there may be other reasonable
choices. For example, if the susceptible had a probability $\lambda$
of contagion with each infected neighbor, then the probability of
infection is $[1-(1-\lambda)^{k_{inf}}]$. For small $\lambda$, we
have $p_i(t)=[1-(1-\lambda)^{k_{inf}}]\approx \lambda*k_{inf}\propto
n_{inf}(t)$. Obviously, that dose not affect the qualitative results
here.

To quantify the amplitudes of the oscillation series, we define the
relevant order parameter
\begin{equation}
\sigma (t)=\left|\frac{1}{N}\sum_{j=1}^N e^{i\,\phi_j(t)}\right|,
\label{eq.4}
\end{equation}
where $\phi_j=2\pi (\tau_j-1) /T$ is a geometrical phase
corresponding to $\tau_j$. The states $\tau=0$ have been left out
of the sum in Eq.(\ref{eq.4}). We obtain the synchronization order
parameter $\sigma$ by averaging over $10^4$ time steps after the
transient to a stationary state and subsequently by averaging over
400 different realizations of the system.
 Here the synchronization is a measure of the collective order.
 If at any time $t$ all the vertices are almost at the same state, i.e.
 $\tau_i$ is equal to the same value for all $i$, the system is synchronous.
 While the vertices are at different states equally, the system is not
 synchronous. Obviously, when the system is not synchronized,
 the phases are widely spread in the cycle and the
complex numbers $e^{i\phi}$ are correspondingly spread in the unit
circle which leads to low value of $\sigma$. In contrast, when a
significant part of the vertices are synchronized in the cycle,
$\sigma$ is large. The full synchronization, i.e. $\sigma=1$, will
be achieved only when all the vertices enter the same state
simultaneously. For the local synchronization, we calculate the
above order parameter over the vertices in one community.

Precise calculation of the critical community strength $Q_c$
separating synchronized and desynchronized states requires
considering the finite-size effect. In the thermodynamic limit, the
order parameter displays the critical behavior
$\sigma\sim(Q-Q_c)^\beta$, with the critical exponent $\beta$. While
in a finite system with size much larger than the additional length
scale, the critical scaling from is
\begin{equation}
\sigma=N^{-\beta/\overline{\nu}}F[(Q-Q_c)N^{1/\overline{\nu}}],
\label{eq.5}
\end{equation}
where the exponent $\nu$ describes the divergence of correlation
volume $\xi_v$ at $Q_c$, $\xi_v\sim|Q-Q_c|^{-\overline{\nu}}$.
Since at $Q=Q_c$ the function $F$ in Eq. (\ref{eq.5}) has a value
independent of $N$, plotting $\sigma N^{\beta/\overline{\nu}}$ vs
$Q$ for various sizes, one can get the value of
$\beta/\overline{\nu}$ that gives a unique crossing point at
$Q_c$. One then use
\begin{equation}
\ln[\frac{d\sigma}{dQ}]_{Q_c}=\frac{1-\beta}{\overline{\nu}}\ln{N}+const
\label{eq.6}
\end{equation}
in order to determine the value of $(1-\beta)/\overline{\nu}$.
Then the exponent $\beta$ and $\overline{\nu}$ can be figured out.
Figure 3(a) displays the determination of $Q_c$ for the global
synchronization using the finite-scale form Eq. (5). Varying the
value of $\beta/\overline{\nu}$ we find that
$\beta/\overline{\nu}\simeq0.25$ gives a well-defined crossing
point at $Q_c\simeq0.835$. In Fig. 3(b), the least-square fit to
Eq. (6) gives $(1-\beta)/\overline{\nu}\simeq0.572$. These yield
$\beta\approx0.30$ and $\overline{\nu}\approx1.22$. Moreover, as
showed in Fig. 3(c), for different mean degree $\langle k\rangle$,
the local synchronization parameter falls into the pit around a
value $Q\approx0.75$. This implicates that, when the communities
are almost unattached (i.e. for very large $Q$) the local dynamic
lies on the inner structure of community independently, while the
communities couple each other strongly (for small $Q$), the local
dynamic is almost the same as the global one, and in the midst the
local synchronization is the weakest.

Further more, we have studied the impact of the mean degree
$\langle k\rangle$ of scale-free networks upon the
synchronization. We start the simulation with a generation of
scale-free networks with $Q=0$, that is the BA model or our model
with $n/m=1/c$, and the initial fraction of infected vertices
$n_\text{inf}(0)=0.1$. We let the number of vertices $m$ that a
new added vertex will connect be real number, as referred in
\emph{section II}. Obviously, $\langle k\rangle=2m$. Fig. 4(a)
shows the order parameter $\sigma$ vs $m$ for different period $T$
of the infection cycle. For a fixed period $T$, a transition in
the synchronization can be observed as $m$ increases. Moreover,
the larger the period T, the less the critical value of mc, at
which the transition occurs. We set $T=10$ to analyze the critical
scaling by using standard finite-size analysis mentioned above.
Fig. 4(b) displays that when $\beta/\overline{\nu}=0.16$ the
curves with different sizes $N$ give a unique crossing point at
$m_c\approx2.8$ (where $\langle k\rangle_c\approx5.6$). In Fig.
4(c), the fit gives $(1-\beta)/\overline{\nu}\simeq0.35$. Hence
$\beta\simeq 0.31$ and $\nu\simeq 1.95$.

\emph{V. Conclusion.}---To summarize, we have investigated the
influence of the strength of community structure (Q) on global and
local synchronization induced by the SIRS epidemic dynamics. The
numerical results have shown that small Q induces better global
synchronization and a phase transition occurs at $Q_c\approx0.835$
estimated by using finite size analysis; While for the local
synchronization there exists a minimal value of order parameter
$\sigma$ around $Q\approx0.75$. This result is in accordance with
Ref. \cite{Entangled} in which a modified simulated annealing
algorithm is applied to optimize the synchronizability and
well-defined communities do not exit in the emerging networks. That
implies the networks with small $Q$ are of strong synchronizability.
It is also worth mentioning that, as in synchronization process
well-defined communities of nodes emerge in different time scales,
Arenas \textit{et al} have used the synchronization to reveal the
community structure \cite{Reveals}.

Moreover, we have studied the synchronization order parameter vs
$\langle k\rangle$ on scale-free networks with $Q=0$. The simulation
results demonstrate that, for a fixed period $T$, a transition in
the synchronization can be observed as $\langle k\rangle$ increases.
The larger period $T$ corresponds to smaller critical value of
transition point $\langle k\rangle_c$.

We acknowledge the support from the National Natural Science
Foundation of China under Grants No. 71471033 and No. 70671097.

\end{document}